\documentclass[toc]{PoS}

\usepackage{amsmath}
\usepackage{amssymb,amscd}
\usepackage[arrow,matrix,curve]{xy}

\usepackage{epsfig}

\def\={\ =\ }
\def\dd{{\rm d}}

\newcommand{\tr}[1]{\:{\rm tr}\,#1}

\DeclareMathOperator{\End}{End}

\def\ii{{\,{\rm i}\,}}

\newcommand{\bbr}{\mathbb{R}}
\newcommand{\bbR}{\mathbb{R}}
\newcommand{\bbc}{\mathbb{C}}

\newcommand{\bbP}{\mathbb{P}}

\newcommand{\CT}{{\rm{CT}}}

\def\alg{{\mathcal A}}

\def\hil{{\mathcal H}}

\def\CCH{{\mathcal H}}
\def\CT{{\mathcal T}}
\newcommand{\RZ}{\mathbb{Z}}
\def\xd{\dot{x}}

\def\Id{{\rm id}}

\newcommand{\bbz}{{\mathbb Z}}

\newcommand{\unit}{\widehat{1}}   			
\newcommand{\id}{\mathrm{id}}   			

\newcommand{\CL}{\mathcal{L}}

\newcommand{\frg}{\mathfrak{g}}				

\newcommand{\ah}{\hat{a}}

\newcommand{\fh}{\hat{f}}

\newcommand{\xh}{\hat{x}}
\newcommand{\di}{\mathrm{i}}     			
\newcommand{\eps}{{\varepsilon}}			

\newcommand{\bz}{{\bar{z}}}

\newcommand{\dderr}[2]{\frac{\dd #1}{\dd #2}}   	

\newcommand{\remark}[1]{}     				
\title{Branes, Quantization and Fuzzy Spheres}

\ShortTitle{Branes, Quantization and Fuzzy Spheres}

\author{\speaker{Christian S\"amann \ and \ Richard J. Szabo} \
         \thanks{Report numbers: \ HWM--11--1 \ , \ EMPG--11--03}
        \\ Heriot--Watt University, Edinburgh, U.K. \\
        E-mail: \email{C.Saemann@hw.ac.uk , R.J.Szabo@hw.ac.uk}}

\abstract{
We propose generalized quantization axioms for Nambu--Poisson manifolds, which allow for a geometric interpretation of $n$-Lie algebras and their enveloping algebras. We illustrate these axioms by describing extensions of Berezin--Toeplitz quantization to produce various examples of quantum spaces of relevance to the dynamics of M-branes, such as fuzzy spheres in diverse dimensions. We briefly describe preliminary steps towards making the notion of quantized 2-plectic manifolds rigorous by extending the groupoid approach to quantization of symplectic manifolds.
}

\FullConference{Corfu Summer Institute on Elementary Particles and Physics -- Workshop on Noncommutative Field Theory and Gravity\\
		 September 8--12, 2010\\
		 Corfu, Greece}
\begin{document}

\section{Brane dynamics and fuzzy spheres}

\subsection{Generalized Nahm equations}

A large variety of quantum field theories arise in the low energy limits of string theory. Many properties and phenomena of string theory are reflected in these low energy descriptions. At the same time, embedding field theories in string theory provides interesting new points of view, leading to unexpected insights and conjectures. It also provides a means of realizing solitonic objects of string theory, collectively called ``branes'', in terms of conventional topological solitons of field theory, such as monopoles, vortices, and instantons. The dynamics of these objects describe the nonperturbative regime of string theory.

For example, the low energy effective field theory on a collection of coincident D3-branes in Type~IIB string theory is $\mathcal{N}=4$ supersymmetric Yang--Mills theory in four dimensions, obtained using T-duality as the dimensional reduction of ten-dimensional $\mathcal{N}=1$ supersymmetric Yang--Mills theory. A class of BPS solutions to the supersymmetric Yang--Mills equations are magnetic monopoles which obey the Bogomol'nyi equations
$$
*F\=D\phi
$$
with Higgs field $\phi$. These monopoles can be interpreted as D1-branes ending on the D3-branes.
Since these objects are BPS configurations, one can trivially form a stack of $N$ D1-branes. From the perspective of the D1-branes, the effective dynamics is described by the {\it Nahm equations}~\cite{Nahm,Diaconescu97,Tsimpis98}
$$
\dderr{X^i}{s}+\eps^{ijk}\,[X^j,X^k]\=0 \ ,
$$
where $X^i(s)$, $i=1,2,3$ are $N\times N$ matrix fields describing the transverse fluctuations inside the D3-brane, and $s$ denotes the zero-mode of the Higgs field $\phi$ which can be identified with the worldsheet spatial coordinate of the D1-branes~\cite{Diaconescu97}. These equations have a ``fuzzy funnel'' solution given by 
$$
X^i(s)\=\frac1s\,\tau^i \qquad \mbox{with} \qquad
\tau^i\=\eps^{ijk}\,[\tau^j,\tau^k]
$$
in terms of generators $\tau^i$ of the Lie algebra
$\mathfrak{su}(2)$. Points in the worldsheet of the D1-brane are thus
blown up into fuzzy two-spheres $S^2$ with radius $\frac1s$. This describes the polarization of the D1-branes which carry magnetic charge due to the anomalous coupling of the D3-branes to a Ramond--Ramond two-form gauge potential~\cite{Myers99}.

While there is an effective description of D-branes, a complete such
theory is lacking for M-branes. The situation just described has a
natural lift to M-theory wherein M2-branes ending on M5-branes are
conjecturally described by Nahm-type equations. The analog of BPS
equations on a flat M5-brane are given by
$$
\partial^2 X^a\=0 \qquad \mbox{and} \qquad \partial^i H_{ijk}\=0 \ ,
$$
where $X^a$ are coordinates along the membranes and $H$ is the three-form field strength on the worldvolume of the M5-brane. For a soliton
solution we take $$H_{01i}\=\partial_i \phi \qquad \mbox{and} \qquad
H_{ijk}\=\eps_{ijkl}\,\partial_l\phi \ , $$ which describes the M2-brane as a solitonic configuration inside the M5-brane. From the perspective of the M2-branes, one postulates the existence of four scalar fields $X^i$, $i=1,2,3,4$, representing the transverse fluctuations inside the M5-brane, and satisfying the {\it Basu--Harvey equations}~\cite{BasuHarvey05}
$$
 \dderr{X^i}{s}+\eps^{ijkl}\, [X^j,X^k,X^l]\=0 \ ,
$$
where the 3-bracket $[-,-,-]$ describes a generalization of the notion of Lie algebra to an entity called a 3-Lie algebra. Similarly to the D1--D3 system, these equations have a solution
$$
X^i(s)\=\frac{1}{\sqrt{2s}} \,\tau^i \qquad \mbox{with} \qquad \tau^i\=\eps^{ijkl}\,[\tau^j,\tau^k,\tau^l] \ .
$$ 
We interpret this solution again as a ``fuzzy funnel'', but this time
with ``fuzzy three-spheres $S^3$'' of radius $\frac1{\sqrt{2s}}$.

One of the goals in the following will be to provide a geometrical meaning and proper definition of these noncommutative spheres. We will work from the fact that generalized Nahm equations are built on \emph{$n$-Lie algebras}. We now explain this concept in more detail.

\subsection{$n$-Lie algebras}

An \emph{$n$-Lie algebra} is a complex vector space $A$ equipped with
a totally antisymmetric $n$-ary map
$$
[-,\dots,-]\,:\, A^{\wedge n}\
  \longrightarrow\ A
$$ 
satisfying the {\it
    fundamental identity}
\begin{eqnarray*}
\big[a^1,\dots ,a^{n-1},[b^1,\dots ,b^n]\big]&=& \sum_{i=1}^n\,
\big[b^1,\dots,b^{i-1}, [a^1,\dots
,a^{n-1},b^i],b^{i+1},\dots ,b^n \big]
\end{eqnarray*}
for $a^i,b^j\in A$, which can be regarded as an ``$n$-Jacobi
  identity''~\cite{Filippov85}. Gauge transformations
  are built from inner derivations
  $\delta:A^{\wedge(n-1)}\longrightarrow {\rm Der}(A)$ defined for $a^i\in A$ via
$$
\delta_{a^1\wedge\dots\wedge a^{n-1}}(b) \ :=\
  [a^1,\dots,a^{n-1},b] \ , \qquad b\in A \ .
$$
The fundamental identity ensures that they form a Lie subalgebra
$\frg_A$ of ${\rm End}(A)$. Note that fixing one slot of an $n$-Lie bracket reduces it to an $(n-1)$-Lie bracket. The case $n=2$ corresponds to the usual
notion of a Lie algebra.

There is a close connection between the $n$-Lie algebras appearing in
brane models and strong homotopy Lie algebras~\cite{Lazaroiu09}. One
can combine the pair of algebras $(A\ ,\ \frg_A)$ into a single space
$\CL$. The fundamental identity and the Jacobi identity are then part of a
chain of homotopy Jacobi identities. Hence $\CL$ is an (ungraded)
\emph{$L_\infty$-algebra} (or ``strong homotopy Lie
  algebra''). In this setting, the generalized Nahm equations are precisely the homotopy
  Maurer--Cartan equation for the $L_\infty$-algebra
  $\CL\otimes\Omega^\bullet(\bbR)$. These $L_\infty$-structures may
  play an important role in the quantization of $n$-Lie algebras, as
  we describe later on.

The basic example of a metric $n$-Lie algebra relevant to the quantization
of spheres is denoted $A_{n+1}$. It is
defined as the vector space $\bbc^{n+1}$ with basis $\tau^1,\dots,\tau^{n+1}$ obeying
$$
[\tau^{i_1},\dots, \tau^{i_n}] \= \eps^{i_1\dots i_ni_{n+1}}\,
    \tau^{i_{n+1}} \ .
$$  
This is the unique simple $n$-Lie algebra over $\bbc$. Its associated
Lie algebra is 
$$
\frg_{A_{n+1}}\= \mathfrak{so}(n+1) \ .
$$
The problem of constructing associative universal enveloping algebras $\mathcal{U}(A_{n+1})$ is addressed in~\cite{DeBellis10b} from the point of view of quantization of duals of $n$-Lie algebras. In~\cite{Bremner10}, noncommutative Gr\"obner bases are used to give a monomial basis for $\mathcal{U}(A_{n+1})$ realized as a quotient of the free associative tensor algebra on a basis of $A_{n+1}$.

\subsection{Geometrical meaning of $n$-Lie algebras}

Nambu--Poisson manifolds yield geometric realizations of $n$-Lie
algebras by taking $A$ to be
the algebra of functions. They generalize the usual notion
of Poisson manifold for $n=2$. A
\emph{Nambu--Poisson structure} on a smooth manifold $M$
  is an $n$-Lie algebra structure 
$$
\{-,\dots,-\} \,:\,
  C^\infty(M)^{\wedge n}\ \longrightarrow\ C^\infty(M)
$$ 
satisfying, in addition to the fundamental identity,
the generalized Leibniz rule
\begin{eqnarray*}
\{f_1 \,f_2,f_3,\dots ,f_{n+1}\}\=f_1\,\{f_2,\dots
 ,f_{n+1}\}+\{f_1,\dots ,f_{n+1}\} \,f_2
\end{eqnarray*}
for $f_i\in C^\infty(M)$.

The basic example of a Nambu--Poisson manifold is the unit $n$-sphere
$S^n$, regarded as an embedded submanifold of Euclidean space $\bbr^{n+1}$ with Cartesian
coordinates $x^1,\dots,x^{n+1}$. The Nambu--Poisson bracket is defined as the extension of the $n$-bracket
$$
\{x^{i_1},\dots, x^{i_n}\}\= \eps^{i_1\dots i_ni_{n+1}}\,
    x^{i_{n+1}}
$$ 
by linearity and the generalized Leibniz rule. Thus $S^n$ provides a
geometric realization of the $n$-Lie algebra $A_{n+1}$. In particular,
the associated Lie algebra $\frg_{A_{n+1}}\=\mathfrak{so}(n+1)$
generates isometries of $S^n$. Thus we may regard fuzzy $n$-spheres as
the appropriate quantization (to be discussed below) of either the
dual of the
$n$-Lie algebra $A_{n+1}$ or of the canonical Nambu--Poisson structure
on $S^n$. For $n=2$ both these quantizations are well-known and
equivalent.

\subsection{Quantum geometry of branes}

We have seen that the effective geometry of the D1--D3 system in
string theory is that of a fuzzy two-sphere $S^2$; this is the most
studied and best understood fuzzy space. 
The question of a low-energy description of stacks of M2-branes has
been the focus of much work in M-theory over the past few years, and
has culminated in the Bagger--Lambert--Gustavsson (BLG) theory of
multiple M2-branes~\cite{Bagger08,Gustavsson09}. In
this article we explore the question of what is the appropriate notion
of noncommutative or deformed geometry that appears in the effective
description of M2-branes and M5-branes, in the context of how the
D-brane realizations of monopoles lift to M-theory. The answers to
these questions might provide deep insight into the nature of M-branes
and could significantly improve the current understanding of
M-theory. In particular, we seek a proper definition of the fuzzy
three-sphere $S^3$ that appears in the M2--M5 system. It is expected
that the quantization of the three-sphere will give interesting hints
on how to appropriately amend the current effective descriptions of
multiple M2-branes. Moreover, by considering stacks of M2-branes
ending on stacks of M5-branes, the properly defined fuzzy $S^3$
should also help in pinning down the gauge structure in an effective
description of stacks of multiple M5-branes.

Another context in which theories on quantized spaces arise in
low-energy effective descriptions is through modifications to
generalized Nahm equations when branes are subjected to
background supergravity fields. In~\cite{ChuSmith09} it is shown how
to interpret the Nahm equations for the D1--D3 system as boundary conditions for open
strings; they are BPS equations in the effective field theory on $N$ coincident D1-branes obtained as the two-dimensional reduction of $\mathcal{N}=1$ supersymmetric Yang--Mills theory with gauge group $U(N)$ in ten dimensions. In particular, a constant $B$-field on the D3-brane induces a
  shift in the Nahm equations, which is accounted for by writing a
  Heisenberg algebra relation
$$
[X^i,X^j]\= \ii\theta^{ij}\ \unit
$$
on the boundary string fields, with central element $\unit$ and noncommutativity bivector 
$\theta\= \frac12\,\theta^{ij}\, \partial_i\wedge \partial_j$ a function of the two-form $B$-field. The effective noncommutative
geometry on the D3-brane in this case is the well-known Moyal space (see e.g.~\cite{Szabo03}), which in the
present context arises via quantization of the constant Poisson bracket on $\bbr^2$~\cite{Ardalan99,ChuHo99,Schomerus99,SeibergWitten99}.

In an analogous way, the Basu--Harvey equations may be regarded as boundary conditions of open
  membranes. In this case the modification due to a constant $C$-field
  on the M5-brane is accounted
  for by postulating the relations of the {\it Nambu--Heisenberg algebra}~\cite{Nambu73}
$$
[X^i,X^j,X^k]\= \ii\Theta^{ijk} \ \unit \ , 
$$
with 3-central element $\unit$ and noncommutativity 3-vector $\Theta\=\frac1{3!}\, \Theta^{ijk}\, \partial_i\wedge\partial_j\wedge\partial_k$ a function of the three-form
$C$-field. In~\cite{ChuSmith09} it is suggested that the associated
noncommutative geometry should arise from quantization of the
constant Nambu--Poisson 3-bracket on $\bbr^3$. In the following we
will focus on the geometries mentioned above, and discuss how to make
sense of both fuzzy $S^3$ and the geometry of the Nambu--Heisenberg
algebra following for a large part the treatment of~\cite{DeBellis10b}. These and other quantized Nambu--Poisson manifolds have
recently been derived as supersymmetric solutions in a 3-Lie algebra
reduced model derived from dimensional reduction of the BLG
model~\cite{DeBellis10a}, which contains the IKKT matrix model as a
strong coupling limit.

\section{Quantization}

\subsection{Axioms of quantization}

The problem of quantization is highly non-trivial and far from being
fully understood. Quantization is essentially a process that
turns ``classical'' objects into their ``quantum'' analogs. At the
classical level, states are points on a Poisson manifold $M$, and
observables are functions on $M$. At the quantum level, states are
rays in a complex Hilbert space $\hil$, and observables are operators
on $\hil$. Thus quantization provides a map
$$
C^\infty(M) \ \longrightarrow \ \End(\hil) \ , \qquad f \ \longmapsto
\ \hat f
$$
that we should subject to a list of axioms depending on the specific
applications. We could also demand that Poisson diffeomorphisms of $M$
map to automorphisms of the associative operator algebra of quantum
observables. If in addition $M$ is a symplectic manifold, then we
should further map Lagrangian submanifolds $L\subset M$ to vectors
$\psi_L\in\hil$.

Dirac's ``wish list'' was that of a full quantization, which is a map satisfying:
\begin{itemize}
\item[1.] The assignment $f \longmapsto \hat f$ is $\bbc$-linear, and
  if $f\= f^*$ is real then $\hat f\= \hat f^\dag$ is Hermitian;
\item[2.] The identity function $f\=1$ is mapped to the identity operator $\id_\hil$;
\item[3.] {Correspondence principle:} \ $[\hat f,\hat g]\=
  -\ii \hbar\, \widehat{\{f,g\}}$ for $f,g\in C^\infty(M)$; and
\item[4.] The quantized coordinate functions $\hat x^i$
  act irreducibly on $\hil$.
\end{itemize}
When dealing with homogeneous spaces $M\=
  G/H$, we may also wish to add:
\begin{itemize}
\item[4a.] The Hilbert space $\hil$ carries a representation of the
  isometry group $G$.
\end{itemize}
The problem with this ``wish list'' is that full quantizations in
general do not exist; in particular, by the Groenewold-van~Howe
theorem there is no full quantization for the symplectic manifolds
$T^*\bbr^n$ and $S^2$.

There are various loopholes to the obstructions to full
quantizations. We shall use Berezin quantization which is a hybrid of
geometric and deformation quantization, and constructs fuzzy
geometry. It utilizes three weakenings of the axioms for
quantization. Firstly, one drops the irreducibility assumption, which
amounts to a \emph{prequantization}; this involves a choice of line
bundle over $M$, which can lead to a quantization of $\hbar^{-1}$
(i.e. a restriction of the deformation parameter $\hbar$ to a discrete subset of
$\bbc^\times$). Secondly, one quantizes only a subset
$\Sigma\subsetneq C^\infty(M)$ of ``quantizable functions'' on $M$; this amounts to choosing a
\emph{polarization}, that corresponds on local charts to representing
$M$ as a cotangent bundle $T^*U$, and leads into \emph{geometric
  quantization}. Finally, one invokes the correspondence principle
only to linear order in $\hbar$; this is the earmark of \emph{deformation quantization}.

\subsection{Axioms of generalized quantization}

In~\cite{DeBellis10b} we proposed a generalization of the quantization
axioms to Nambu--Poisson manifolds. This problem is notoriously
difficult, and many previous efforts were devoted to extending
geometric quantization (see~\cite{DeBellis10b} for a list of references). Here we will extend Berezin quantization to
Nambu--Poisson manifolds, keeping the data of a complex Hilbert space
$\hil$ and $\End(\hil)$ as the algebra of quantum observables. The
extension requires the formulation of generalized quantization axioms,
that we now describe.

A Nambu--Poisson quantization is a map
$Q:\Sigma\longrightarrow\End(\hil)$, $\Sigma\subsetneq C^\infty(M)$,
  satisfying:
\begin{itemize}
\item[1$'$.] The assignment $f \longmapsto \hat f\=Q(f)$ is $\bbc$-linear, and
  if $f\= f^*$ is real then $\hat f\= \hat f^\dag$ is Hermitian;
\item[2$'$.] The identity function $f\=1$ is mapped to the identity
  operator $\id_\hil$; and
\item[3$'$.] {Correspondence principle:} 
$$
\lim_{\hbar\rightarrow 0}\,\Big\|
\frac{\di}{\hbar}\,\sigma\big([\fh_1,\ldots,\fh_n]\big)-\{f_1,\ldots,f_n\}\Big\|_{L^2}
\=0
$$
\end{itemize}
for $f_i\in C^\infty(M)$, where $\sigma:Q(\Sigma)\longrightarrow\Sigma$ is the symbol map and the $L^2$-norm is taken with respect to a chosen measure on $M$. If $M$ is a Poisson manifold, then these axioms are always
satisfied for Berezin quantization.

Axiom 3$'$ requires the choice of an $n$-Lie bracket on $\End(\hil)$. A
natural choice can often be obtained by {\it truncating} the Nambu--Poisson algebra on
the algebra of polynomials $\bbc[x^i]$ to obtain a corresponding
$n$-Lie algebra~\cite{HoHou08}. Denote the truncated Nambu--Poisson
bracket to polynomials of degree $\leq K$ by $\{-,\dots,-\}_K$, and introduce
$$
[\fh_1,\ldots,\fh_n] \ :=\ \sigma^{-1}\big(-\di\, \hbar\,
\{\sigma(\fh_1),\ldots,\sigma(\fh_n)\}_K \big)
$$
with $K\longrightarrow \infty$ in the classical limit $\hbar \longrightarrow
0$. Then the correspondence principle always holds automatically. This
$n$-bracket is in general a deformation of the totally antisymmetric operator product $\eps^{i_1\cdots i_n}\, \fh_{i_1}\cdots\fh_{i_n}$.

\section{Berezin--Toeplitz quantization}

We now review the basic aspects of Berezin quantization that we shall need, particularly of complex
projective space. Generally, on a K\"ahler manifold $(M,\omega)$, we
take as Hilbert space the vector space $$\hil_L\= H^0(M,L)$$ of global holomorphic
  sections of a ``quantum
  line bundle'' $L$ over $M$ with first Chern class $c_1(L)\=
  [\omega]$; this is the choice of holomorphic polarization. The existence of this holomorphic line bundle imposes the
  quantization condition
$$
[\omega]\ \in\ H^2(M,\bbz) \ ,
$$
and the space $(M,\omega)$ is quantizable only if this constraint is met.

For $M\=\bbc\bbP^n$, we take the natural K\"ahler two-form $\omega$ corresponding to the Fubini--Study metric, and set $L \ :=\ \mathcal{O}(k)$. Then the finite-dimensional Hilbert space $\hil_k\ :=\ \hil_L$ consists of homogeneous polynomials of degree $k$ in the homogeneous coordinates $z_0,z_1,\dots,z_n$ of $\bbc\bbP^n$ and can be presented as
$$
\hil_k\={\rm span}_\bbc(z_{\alpha_1} \cdots
z_{\alpha_k})_{\alpha_i=0}^n\= {\rm span}_\bbc\big(
\ah^\dagger_{\alpha_1}\cdots \ah^\dagger_{\alpha_k}|0\rangle \big) \ ,
$$
where $\ah_\alpha,\ah^\dag_\alpha$ are the standard creation and annihilation operators of an $(n+1)$-dimensional harmonic oscillator. This space coincides with the
$k$-particle Hilbert space of ``lowest Landau level states'' in the generalization of the quantum Hall effect (Landau problem) on $\bbc\bbP^n$~\cite{Karabali06}.

Generally, an overcomplete basis of the Hilbert space is given by the {Rawnsley coherent states}. Associated to any $z\in M$ there is a corresponding coherent state $|z\rangle \in\hil_L$. For $M\=\bbc\bbP^n$, the states 
$$
|z\rangle\=\frac{1}{k!}\, \big(\bz_\alpha \,
  \ah_\alpha^\dagger \big)^k|0\rangle
$$
are the usual Perelomov coherent states. The quantization is now set up by using the coherent states as a bridge between classical and quantum observables in two ways through the Berezin symbol and quantization maps
$$\displaystyle{ f(z)\=\sigma\big(\hat{f}\,
    \big)\=\frac{\langle z|\hat{f}\, |z\rangle}{\langle z | z \rangle}
    \qquad \mbox{and} \qquad \hat{f}\= Q(f)\= \int_M \, \frac{\omega^{n}}{n!}~
    \frac{|z\rangle\langle z|}{\langle z | z \rangle}\, f} \ . $$ 
For $M\=\bbc\bbP^n$ this quantization map defines the fuzzy projective space $\bbc\bbP^n$. It obeys the convergence property
$$\displaystyle{\lim_{k\to \infty} \, \Big\| \ii
  k\,\big[Q(f)\,,\,Q(g)\big]- Q\big(\{f,g\}\big)\Big\|_{\rm HS}\=
  0}$$ 
with respect to the Hilbert--Schmidt norm on $\End(\hil_k)$~\cite{Bordemann94}.

The Rawnsley coherent states also have geometrical applications. The Bergman metric for K\"ahler manifolds is given by $$g\= \frac1k\,
  \partial\,\overline\partial\,\log\langle z | z \rangle \ .$$ 
The expansion of this metric in the classical limit $k\longrightarrow \infty$ in powers of curvatures
approximates Einstein metrics for projective K\"ahler manifolds, e.g. for Calabi--Yau manifolds embedded in $\bbc\bbP^n$. To leading orders one has~\cite{Zelditch98}
$$
\langle z | z \rangle\= \omega^{n}+\omega^{n-1} \ \frac R2+\cdots \ .
$$
For holomorphically embedded submanifolds $M\subset\bbc\bbP^n$, given by the zero locus
$$
M\=\big\{z\in\bbc\bbP^n\ \big|\ p(z)=0\big\}
$$
of a holomorphic function $p(z)$ in the homogeneous coordinates of $\bbc\bbP^n$, one can factor the algebra of functions on $\bbc\bbP^n$ by the corresponding ideal and quantize $M$ on the Hilbert space~\cite{Saemann08}
$$
\hil_M=\big\{|\mu\rangle\in \hil_L \ \big|\ \hat p|\mu\rangle=0\big\} \ .
$$

\section{Quantization of spheres}

\subsection{Quantization of $S^2$: \ The fuzzy sphere}

The Berezin quantization of $\bbc\bbP^1$ is the celebrated fuzzy sphere~\cite{Berezin75,Madore92}. For $\bbc\bbP^1 \ \cong\ S^2$, the space of quantizable functions 
  $\Sigma$ is spanned by spherical harmonics $Y_{l,m}$ with $l\leq k$.
The Poisson bracket $$\{x^i,x^j\}\= \eps^{ijk}\, x^k$$ on $S^2$ maps to the $\mathfrak{su}(2)$ Lie algebra $$[\hat x^i,\hat x^j]\= -\ii \hbar\,
  \eps^{ijk}\, \hat x^k$$ under the correspondence principle. 

The quantization map is described by the Jordan--Schwinger transformation which sends the local coordinates
$$\displaystyle{x^i\=\frac{\bz_\alpha\, \sigma^i_{\alpha\beta}\,
      z_\beta}{|z|^2} \ \in\ S^2\subset\bbr^3}$$
to the operators
$$
 \xh^i\=\frac{1}{k!}\,\sigma^i_{\alpha\beta}\,\hat a^\dagger_\alpha\,
 \hat a^\dagger_{\rho_1}\cdots \hat
 a^\dagger_{\rho_{k-1}}|0\rangle\langle 0|\hat a_\beta\, \hat
 a_{\rho_1}\cdots \hat a_{\rho_{k-1}} \ . $$ 
A straightforward calculation gives $$\displaystyle{\hbar\=\frac2k}$$ in this case, so that the classical limit is
  $k \longrightarrow \infty$.
This construction generalizes to any projective space $\bbc\bbP^n$ by replacing the $SU(2)$ Pauli spin matrices 
  $\sigma^i_{\alpha\beta}$ with the Gell-Mann matrices
  $\lambda^i_{\alpha\beta}$ of its isometry group $SU(n+1)$. The
  corresponding (non-formal) coherent state star-product $f\star g\=\sigma(\hat f\, \hat g)$ is computed explicitly in~\cite{Balachandran02}.

\subsection{Quantization of $S^4$}

Our quantization of $S^4$ yields the noncommutative spheres of Guralnik and Ramgoolam~\cite{Guralnik01}. For this, we use the Clifford algebra $Cl(\bbr^5)$ to isometrically embed $S^4\subset\bbc\bbP^3$ via the map
$$
 x^i\=\frac{1}{|z|^2}\, \gamma^i_{\alpha\beta}\, \bz^\alpha\, z^\beta \qquad \mbox{with} \qquad x^i\, x^i\= 1 \ .
$$
This embedding is not holomorphic. An alternative quantization in the same spirit is obtained by considering the sphere fibration $\bbc\bbP^3 \xrightarrow{S^2} S^4$~\cite{Medina03,Dolan03,Abe04}.

The restricted coherent state projection of the Berezin symbol $\sigma(\fh)$, $\fh\in\hil_k$
  to $\Sigma\cap C^\infty(S^4)$ gives the quantization map
$$
\hat{x}^i \ :=\ \frac{1}{k!}\, \gamma^i_{\alpha\beta}\,
\ah^\dagger_\alpha\, \ah^\dagger_{\rho_1} \cdots
  \ah^\dagger_{\rho_{k-1}}|0\rangle\langle 0| \ah_\beta\,
  \ah_{\rho_1}\cdots \ah_{\rho_{k-1}} \ ,
$$
and the endomorphism algebra
$\End(\hil_k)$ consists of the noncommutative polynomials of degree~$3k$.
This map yields the quadratic Casimir eigenvalue $$\hat{x}^i \,\hat{x}^i\=\Big(1+\frac4k\Big)\
  \id_{\hil_k} \ , $$ and the 4-Lie bracket is identical to the totally antisymmetric
  operator product at affine level. In this case the correspondence principle is satisfied with the truncated
  Nambu--Poisson 4-bracket $\{-,\dots,-\}_k$.

\subsection{Quantization of $S^3$}

To quantize the three-sphere, we consider the surjection $S^4 \longrightarrow S^3$ given by $x^5\=0$. This induces an embedding $S^3\subset\bbc\bbP^1\times\bbc\bbP^1
  \subset\bbc\bbP^3$. However, the constraint
$x^5\=0$ is not holomorphic, and so we cannot factorize the operator algebra $\End(\hil_k)$ by a
  holomorphic ideal. Thus we must project the Hilbert space $\hil_k$ onto a maximal set of irreducible representations of the isometry group $SO(5)$ on
  which the Casimir operator $\hat{x}^i \,\hat{x}^i$
is proportional to the identity operator $\id_{\hil_k}$. This yields a nonassociative operator algebra~\cite{Guralnik01}.

The corresponding 3-Lie algebra is defined by~\cite{BasuHarvey05}
$$
[\hat x^i,\hat x^j,\hat x^k] \ :=\ -[\hat x^i,\hat x^j,\hat x^k,\hat
x^5] \= \ii \hbar(k)\, \eps^{ijkl}\, \hat x^l \ .
$$
In previous works, the additional radial fuzziness in the normal directions to $S^3$ in $\bbc\bbP^1\times\bbc\bbP^1
  \subset\bbc\bbP^3$ was dealt with appropriately; either the radial modes are projected out after operator
  multiplication~\cite{Grosse96,Guralnik01} or are dynamically suppressed~\cite{Medina03}. In our case, we choose to keep these modes. Then the radial fuzziness of our quantum $S^3$ allows for consistent solutions to the Basu--Harvey equations~\cite{Nastase09}.

\subsection{Fuzzy scalar field theory on $S^4$}

In addition to their relevance in the low-energy descriptions of brane dynamics, quantum field theories on Berezin-quantized spaces provide an interesting alternative to lattice regularization. Let us consider the particular example of the noncommutative four-sphere. The Hilbert space $\hil_k\= H^0(\bbc\bbP^3,\mathcal{O}(k))$ carries both an irreducible representation of the
  $SU(4)$ isometry group of $\bbc\bbP^3$ and the spinor representation of the $SO(5)$ isometry group of $S^4$. The Laplace operator $\Delta$ on the algebra of functions $C^\infty(S^4)$ corresponds to the quadratic Casimir
  operator of $SO(5)$ in the spinor representation. It carries over to a linear
  operator on the endomorphism algebra $\End(\hil_k)$ via the ``Berezin push''~\cite{Lazaroiu08}
$$
\Delta^B \ :=\ Q\circ \Delta\circ\sigma \ .
$$
An action functional for scalar fields may now be constructed through the integral 
$$ \displaystyle{
\int_{S^4}\, \dd\mu_{S^4}~f\= \frac1{{\rm vol}(S^2)}\, \int_{\bbc\bbP^3}\,
\frac{\omega^3}{3!}~\rho(f)\= \frac{{\rm vol}(S^4)}{k}\, \tr(\fh)
} \ , $$ 
where $\rho(f)$ is the image of the function $f\in C^\infty(S^4)$ in
$C^\infty(\bbc\bbP^3)$. The corresponding path integral is defined by taking the integration domain to be the space $\Sigma\cap
  C^\infty(S^4)$ of quantizable functions. Thus fuzzy quantum field theories can be used to dynamically distinguish quantum spheres from
 quantum projective spaces.

\section{Quantization of non-compact manifolds}

\subsection{Quantization of $\bbr^3$}

The geometry of the quantized Nambu--Heisenberg algebra $$[\hat
  x^1,\hat x^2,\hat x^3]\= -\ii\hbar \ \unit $$ can be realized in terms of the noncommutative space
  $\bbr_\hbar^3$ which was defined in~\cite{Hammou02}. In this instance there is no truncated
  3-bracket ensuring the correspondence principle. Hence a 3-Lie algebra
  structure is only realized at affine level. One possible model for
  this space is as follows in terms of~$\bbr_\hbar^3$.

Take the fuzzy sphere $S^2$ with Hilbert space $$\hil_k\=
  H^0\big(\bbc\bbP^1\,,\,\mathcal{O}(k) \big) $$ and coordinate generators $\hat x^i$ obeying $[\hat x^i,\hat x^j]\= -\frac{2 \ii}k \,
  \eps^{ijk}\, \hat x^k$. Define
$$
[\xh^1,\xh^2,\xh^3]\=\eps^{ijk}\, \xh^i\, \xh^j\, \xh^k\=-
\frac{6\ii}{k}\ \id_{\hil_k} \ .
$$
The radius $R_k$ of this fuzzy sphere, defined by the quadratic Casimir eigenvalue $\hat x^i\,\hat x^i\= R_k^2\ \id_{\hil_k}$, is given by
$$R_{k}\=\sqrt{1+\frac{2}{k}}~
\sqrt[3]{\frac{\hbar\, k}{6}} \ . $$ 
The Hilbert space and algebra of quantized functions on $\bbr_\hbar^3$
are given by
$$\hil\=\bigoplus_{k=1}^\infty\, \hil_k \qquad \mbox{and} \qquad
\alg\=\bigoplus_{k=1}^\infty \,
  \End(\hil_k) \ . $$ 
This describes the space $\bbr_\hbar^3$  as a ``discrete foliation'' of $\bbr^3$ by fuzzy
spheres~\cite{Hammou02,Batista03} with radius $R_k$. Note that, as expected from the non-compactness of $M$ in this case, there is no quantization condition on $\hbar\in\bbc$ and the Hilbert space $\hil$ is infinite-dimensional.

\subsection{Quantization of hyperboloids}

Our construction can also be extended to the quantization of the
hyperboloids $H^{p,q}$, which can be realized as the non-compact
homogeneous space
$$
H^{p,q}\= SO(p,q)\, \big/ \, SO(p-1,q) \ , $$ 
or alternatively as the quadric 
$$x^\mu\,x^\nu\,
  \eta_{\mu\nu}\=1
$$ 
in $\bbr^{p+q}$ with metric $(\eta_{\mu\nu})$ of signature $(p,q)$. In
this case we must allow for an indefinite metric in the corresponding
Clifford algebra. Hence the quantization map involves non-Hermitian
operators and non-unitary representations. Important examples in
quantum gravity and string theory involve fuzzy $AdS$, and the
quantized M5-brane geometry $
  \bbr^{1,2}_\hbar\times \bbr^3_\hbar$. The associated $n$-Lie algebra
  $A_{p,q}$ is the unique simple $n$-Lie algebra over
  $\bbr$, where $n\=p+q-1$. See~\cite{DeBellis10b} for further details.

\section{Gerbes and quantization of loop spaces}

\subsection{Quantization of 2-plectic manifolds}

We will now sketch some ways in which one may arrive at a complete
quantization of $S^3$ and other Nambu--Poisson manifolds. As we discussed, a symplectic
manifold $(M,\omega)$ is quantizable if it admits a Hermitian line
bundle $L$ with a unitary connection $\nabla$ of curvature
$F_\nabla \=-2\pi\ii\omega$. This two-form is a representative of the
first Chern class $c_1(L)$ which is a characteristic class of the line
bundle $L\longrightarrow M$, and which is an element of the integer cohomology
group $H^2(M,\RZ)$. This class can be constructed explicitly in \v{C}ech cohomology. Given an open cover $(U_i)_{i\in I}$ of the
manifold $M$, we construct the curvature two-form $F_\nabla$, the local one-form gauge potentials $A_{(i)}$
defining the connection $\nabla$ on $U_i$, and the transition functions
$g_{(ij)}=g_{(ji)}^{-1} :U_i\cap U_j\longrightarrow U(1)$ between neighbouring charts which define a \v{C}ech one-cocycle, i.e. they
obey the cocycle condition
$$
g_{(ij)}\, g_{(jk)}\, g_{(ki)}\= 1 \qquad \mbox{on} \qquad U_i\cap
U_j\cap U_k \ .
$$
They are all related through
\begin{eqnarray*}
  F_\nabla &=& \dd A_{(i)} \qquad \mbox{on} \qquad U_i \ , \\[4pt]
A_{(i)}-A_{(j)} &=& \dd \, \log g_{(ij)} \qquad \mbox{on} \qquad
U_i\cap U_j \ .
\end{eqnarray*}
These relationships are obtained directly from repeated applications
of Poincar{\'e}'s lemma.

The Hilbert space $\CCH$ is then constructed by
restricting the space of sections of $L$. For example, we saw that in
Berezin quantization it is given by the vector space $\CCH\=H^0(M,L)$ of global
holomorphic sections of the quantum line bundle $L$. A quantization is completed by specifying a
prescription of how to map a certain subset of functions on $M$ to the
$C^*$-algebra of linear operators on $\CCH$.

Let us attempt to generalize this picture to a 2-plectic manifold
$(M,\varpi)$, again endowed with an open cover $(U_i)_{i\in I}$. Such
a manifold comes with a closed three-form $\varpi$ which gives rise to
a Nambu--Poisson bracket, analogously to the way in which a symplectic
structure gives rise to a Poisson bracket. We thus expect the
quantization condition $\varpi\=\frac{\di}{2\pi}\, H$, where $H$ is a
representative of a class in the integer cohomology group $$[\varpi] \
\in \ H^3(M,\RZ) \ . $$ Just like the first Chern class is a
characteristic class for a Hermitian line bundle with connection, this {\em
  Dixmier--Douady class} is a characteristic class for a bundle
gerbe with connection and curving $H$. In the setting of abelian local gerbes this also has a definition in \v{C}ech cohomology. From the closed three-form $H$, we obtain by repeated
application of Poincar{\'e}'s lemma also a 1-connection, i.e. a family
of two-forms $B_{(i)}$ on patches $U_i$, a
0-connection, i.e. a family of connections $\nabla_{(ij)}\=\nabla_{(ji)}^*$ on line bundles
$L_{(ij)}\= L_{(ji)}^*$ over intersections of patches $U_i\cap U_j$ representing the
first \v{C}ech cohomology, and bundle isomorphisms
$h_{(ijk)}:L_{(ij)}\otimes L_{(jk)}\longrightarrow L_{(ik)}$ on triple
intersections $U_i\cap U_j\cap U_k$ obeying the coherence condition
$$
h_{(ikl)}\circ\big(h_{(ijk)}\otimes \Id_{L_{(kl)}} \big)\= h_{(ijl)}\circ
  \big(\Id_{L_{(ij)}}\otimes h_{(jkl)}\big) \qquad \mbox{on} \qquad
  U_i\cap U_j\cap U_k\cap U_l \ .
$$
They are all related through
\begin{eqnarray*}
H&=&\dd B_{(i)} \qquad \mbox{on} \qquad U_i \ , \\[4pt]
B_{(i)}-B_{(j)}&=& F_{\nabla_{(ij)}} \qquad \mbox{on}\qquad U_i\cap U_j \ ,\\[4pt]
h_{(ijk)}\circ\big(\nabla_{(ij)}\otimes1+1\otimes\nabla_{(jk)} \big) &=&
\nabla_{(ik)}\circ h_{(ijk)} \qquad
\mbox{on}\qquad U_i\cap U_j\cap U_k \ .
\end{eqnarray*}
Note that the bundle isomorphisms $h_{(ijk)}$ are equivalent to the specification of transition functions $g_{(ijk)}:U_i\cap U_j\cap U_k\longrightarrow U(1)$ on triple intersections which define a \v{C}ech two-cocycle such that
\begin{eqnarray*}
A_{(ij)}-A_{(ik)}+A_{(jk)} &=& \dd\, \log g_{(ijk)} \qquad \mbox{on} \qquad U_i\cap U_j\cap U_k \ , \\[4pt]
g_{(ijk)}\, g_{(ijl)}^{-1}\, g_{(ikl)}\, g_{(jkl)}^{-1} &=& 1 \qquad \mbox{on} \qquad U_i\cap U_j\cap U_k\cap U_l \ ,
\end{eqnarray*}
where $F_{\nabla_{(ij)}}\=\dd A_{(ij)}$.

A naive approach to the quantization of a 2-plectic manifold
$(M,\varpi)$ would therefore construct a Hilbert space $\CCH$ from
``global holomorphic sections'' of the gerbe, and the $C^*$-algebra
corresponding to quantized functions from linear operators on $\CCH$; the natural notion of ``polarization'' now appears to be provided by the first cojet bundle rather than the cotangent bundle~\cite{Baez10}. On $M\= S^3\=U_1\cup U_2$ with $U_1\cap U_2\ \cong \ S^2\times
(-1,1)$, this may be attained by pullback from the corresponding
structures on $S^2$ through the Hopf fibration $\pi: S^3\longrightarrow
S^2$; the connection one-form $\kappa$ on this bundle defined via pullback $\dd\kappa\=\pi^*(\omega)$ of the canonical symplectic two-form $\omega$ on $S^2$ defines a contact structure on $S^3$, with volume form $\dd\mu_{S^3}\=\kappa\wedge\dd\kappa$, and hence in this case one quantizes the contact manifold $(S^3,\kappa)$. However, in general all of these objects need to be defined precisely. 

To find a proper quantization of 2-plectic structures, one can
alternatively regard a gerbe as a sheaf of groupoids with locally
isomorphic objects. In this setting one may extend the
reinterpretation of geometric quantization in terms of integration of
Lie algebroids by Hawkins and others; see~\cite{Hawkins06} and
references therein. A \emph{Lie algebroid} over $M$ is a vector bundle $E\longrightarrow M$ together with a Lie bracket on the space of sections $C^\infty(M,E)$ and an ``anchor'' morphism $\rho:E\longrightarrow TM$ to the tangent bundle which implements a Leibniz rule for the bracket; the simplest example is the tangent Lie algebroid with $E\= TM$ and $\rho\=\Id_{TM}$. When $M$ is a point a Lie algebroid is the same thing as a Lie algebra.

In this setting one starts from the observation that the
quantization of the dual of a Lie algebra $\frg$ yields a
$C^*$-algebra which is the convolution algebra of one of the Lie
groups integrating $\frg$. This can be generalized in the following
manner. Every quantizable symplectic manifold comes with a natural Lie
algebroid, the Atiyah Lie algebroid which is an extension of the tangent bundle
$$
{\rm ad}(L) \ \longrightarrow \ TL\,\big/\, U(1)\ \stackrel{\rho}{\longrightarrow} \ TM
\ ,
$$
and which can be integrated to
a Lie groupoid, abstractly a category with smooth structure in which every morphism is invertible. The $C^*$-algebra arising in the quantization of
symplectic manifolds should therefore be identified with the
convolution algebra of the Atiyah Lie groupoid. The connection
$\nabla$ on the line bundle $L$ defines a splitting $s:TM\longrightarrow TL/U(1)$ of this exact sequence, and one can construct a Lie algebra
homomorphism $C^\infty(M)\longrightarrow C^\infty(M,TL/U(1))$ to the
$U(1)$-invariant vector fields on the total space of $L$. In this manner, one can
define polarizations for Lie groupoids, and reconstruct both geometric and Berezin quantization.

In~\cite{Rogers10} it was demonstrated that the Lie algebroid of a
symplectic manifold is replaced by the Courant algebroid on a
2-plectic manifold; a \emph{Courant algebroid} is a vector bundle $E\longrightarrow M$ that generalizes the structure of a Lie algebroid equiped with an inner product on the fibres of $E$. Given a bundle gerbe with connection, on each
patch $U_i$ one builds the standard Courant algebroid $E_i:=TU_i\oplus
T^*U_i$, and then glues together on double intersections using the
0-connection curvatures $F_{\nabla_{(ij)}}$. This gives a Courant algebroid
$E\longrightarrow M$ which is an extension of the tangent bundle
$$
T^*M\ \stackrel{\rho^*}{\longrightarrow}\ E \ \stackrel{\rho}{\longrightarrow} \ TM \ ,
$$
as well as a splitting $s:TM\longrightarrow E$ of this exact sequence
given by the 1-connection $\{B_{(i)}\}_{i\in I}$. The 2-plectic space
$(M,\varpi)$ has a natural $L_\infty$-algebra structure with de~Rham differential $\dd$, which embeds in a canonical $L_\infty$-algebra associated to the
Courant algebroid $E\longrightarrow M$ with differential $\dd_E\=\rho^*\,\dd$. Although the issue of integrating such Courant
algebroids has not been completely settled yet, it is tempting to try to
generalize Hawkins' approach to quantization to this setting. One
advantage of the groupoid construction, besides its generality, is the
fact that it avoids the construction of a Hilbert space and comes
directly to the $C^*$-algebra. For $M\= S^3$, this algebra should represent the proper quantization of the enveloping algebra $\mathcal{U}(A_4)$. Given the difficulties with defining
the notion of global holomorphic sections of a gerbe, this approach
might prove very fruitful.

\subsection{Transgression to loop space}

An alternative approach to the quantization of Nambu--Poisson
structures employs a trick to substitute the gerbe by a principal
$U(1)$-bundle. The price one has to pay is that the base manifold $M$
is replaced by an infinite-dimensional manifold, the loop space of
$M$. Geometric quantization of loop spaces is discussed in the example
$M\= S^3$ in~\cite{Brylinski93}, and more generally in the setting of
infinite-dimensional K\"ahler geometry in~\cite{Sergeev08}. The appearence of the loop space is very natural in the context of the quantization of Nambu--Poisson manifolds. Firstly, this was originally observed in~\cite{Takhtajan94} where an action principle for Nambu mechanics on $\bbr^3$ was formulated as the dynamics of loops; Nambu mechanics has also been recently proposed as a low energy effective description of strings and membranes in constant three-form and four-form flux backgrounds~\cite{ChuHo10}. Secondly, the original attempts~\cite{Bergshoeff00,Kawamoto00} to quantize open membranes ending on an M5-brane in a large constant $C$-field background of M-theory proceeded in analogy to the quantization of open strings ending on a D-brane in a large $B$-field~\cite{SeibergWitten99}. In contrast to the point particle endpoints of open strings, the boundaries of open membranes are closed strings, and this leads to a noncommutative loop space structure. In~\cite{Bergshoeff00,Kawamoto00} canonical quantization of the M5-brane theory was considered with the natural Poisson structure on loop space, leading to a complicated nonassociative algebraic structure. Finally, the map to the loop spaces of $\bbr^4$ and $S^3$ has proved to very helpful in generalizations of the ADHMN construction of monopoles (D1-branes) to the self-dual strings of the six-dimensional theory on an M5-brane~\cite{Gustavsson08,Saemann10}.

This trick is called a {\em transgression} and works as follows. Consider the correspondence
\begin{equation*}
\xymatrix{
 & \CL M\times S^1 \ar[dl]_{ev} \ar[dr]^{\oint_{S^1}} & \\
M &  & \CL M
} 
\end{equation*}
between $M$ and its free loop space $\CL M$. Here $ev$ is the obvious evaluation map of the loop at the given angle in $S^1$, and $\oint_{S^1}$ is the integral over the angle parameterizing the loop. The transgression map $\CT:\Omega^{\bullet+1}(M)\longrightarrow \Omega^\bullet(\CL M)$ amounts to the pullback along $ev$ and the pushforward along $\oint_{S^1}$, i.e. $\CT\=\big(\oint_{S^1}\big)_!\circ ev^*$. Explicitly, it is given by 
\begin{equation*}
 (\CT\alpha)_x\big(v_1(\tau)\,,\,\ldots\,,\,v_k(\tau) \big) \= \oint_{S^1}\, \dd\tau\ \alpha\big(v_1(\tau)\,,\,\ldots\,,\,v_k(\tau),\dot{x}(\tau)\big)~, \qquad \alpha\in\Omega^{k+1}(M)
\end{equation*}
for $x\in M$, where we used the natural tangent vector $\xd(\tau)\in \CL TM\=T \CL M$ available on loop space to fill one of the slots of the $(k+1)$-form $\alpha$ on $M$. 

Using $\CT$, we can thus map the Dixmier--Douady class on a 2-plectic manifold $M$ to a first Chern class on the free loop space $\CL M$. Put differently, a bundle gerbe on $M$ gives rise to a line bundle on $\CL M$. In principle, the latter can then be quantized in the usual way. The difficulty here is that one is working with the infinite-dimensional base space $\CL M$.

Nevertheless, the transgression map can be used to connect the different perspectives on the quantum geometry of M5-branes. The conventional approach, cf.\ e.g.\ \cite{ChuSmith09}, yields a noncommutative space with the description
\begin{equation*}
 \big[x^i\,,\,x^j\,,\,x^k \big]\=\ii\Theta^{ijk}~ \widehat 1 \ ,
\end{equation*}
where $[-,-,-]$ is some 3-algebraic structure, e.g. the bracket of a 3-Lie algebra. Its transgression corresponds to noncommutative loop space relations
\begin{equation*}
\big[x^i(\tau)\,,\,x^j(\sigma) \big]\=\ii\Theta^{ijk}\, \xd^k(\tau)\ \delta(\tau-\sigma)~ \widehat 1 \ .
\end{equation*}
This coincides with the noncommutative loop space structure derived in~\cite{Bergshoeff00,Kawamoto00}.

\section*{Acknowledgments}

RJS would like to thank Dorothea Bahns, Harald Grosse and George Zoupanos for the invitation and kind hospitality at the Corfu Summer Institute. This work was supported by grant ST/G000514/1 ``String Theory
Scotland'' from the UK Science and Technology Facilities Council. The
work of CS was supported by a Career Acceleration Fellowship from the
UK Engineering and Physical Sciences Research Council.

\end{document}